\documentclass[twocolumn,showpacs,preprintnumbers,amsmath,amssymb]{revtex4}
\usepackage{graphicx}
\usepackage{dcolumn}
\usepackage{bm}

\begin{document}
\title{Analysis of $\Lambda_{b} \rightarrow$ $\Lambda \mu^{+} \mu^{-}$ decay in scalar leptoquark model}
\author{Shuai-Wei Wang}
\email{shuaiweiwang@sina.com} \affiliation{ \small Institute of Particle Physics, Central China Normal
University, Wuhan, Hubei  430079, P.~R. China; \\ \small Department of Physics, Nanyang Normal University,
Nanyang, Henan 473061, P.~R. China}
\author{Ya-Dong Yang}\email{yangyd@mail.ccnu.edu.cn}
\affiliation{\small Institute of Particle Physics, Central China Normal
University, Wuhan, Hubei  430079, P.~R. China}

\date{\today}

\begin{abstract}
We analyze the baryonic semilepton decay $\Lambda_{b} \rightarrow$ $\Lambda \mu^{+} \mu^{-}$ in the scalar leptoquark models with $X(3,2,7/6)$ and $X(3,2,1/6)$ states, respectively. We also discuss the effects of this two NP models on some physical observables. For some measured observables, like the differential decay width, the longitudinal polarization of the dilepton system, the lepton-side forward-backward asymmetry and the baryon-side forward-backward asymmetry, we find, the prediction values of SM are consistent with the current data in the most $q^{2}$ ranges, where the prediction values of this two NP models can also keep consistent with the current data with $1\sigma$. However, in some $q^{2}$ ranges, the prediction values of SM are difficult to meet the current data, but the contributions of this two NP models can meet them or keep closer to them. For the double lepton polarization asymmetries, $P_{LT}$, $P_{TL}$, $P_{NN}$ and $P_{TT}$ are sensitive to the scalar leptoquark model $X(3,2,7/6)$ but not to $X(3,2,1/6)$. However, $P_{LN}$, $P_{NL}$, $P_{TN}$ and $P_{NT}$ are not sensitive to this two NP models.
\end{abstract}


\maketitle

\section{\label{sec:level1}Introduction}

The current data have hinted at several anomalies in $B$ decays induced by the flavor-changing neutral current (FCNC) processes $b \rightarrow$ $s \ell^{+} \ell^{-}$, which have been recognized as very important probes of the Standard Model(SM) and new physics (NP). For the baryonic semilepton decays, experimentally, $\Lambda_{b}\rightarrow$ $\Lambda \mu^{+}\mu^{-}$ decay has been observed by the CDF collaboration\cite{aaltonen:2011ss} and measured by the LHCb collaboration at CERN\cite{aaij:2013st}. Theoretically, studies on the $\Lambda_{b}\rightarrow$ $\Lambda \mu^{+}\mu^{-}$ decay have been investigated in the SM and beyond the SM\cite{chenaliev,chenaliev1,chenaliev2}. Their results showed that some observables of these processes are sensitive to the contributions of NP.

The leptoquark models have many kinds of states, not only vector ones, but also scalar ones. In regard to different decay processes, the different leptoquark states may produce different effects. For $b \rightarrow$ $s \mu^{+}\mu^{-}$ processes, model independent constrains on leptoquarks are obtianed in Ref.\cite{nejc1}, where scalar leptoquark states with $X=(3,2,7/6)$ and $X=(3,2,1/6)$ have visible effects on the $b \rightarrow$ $s \mu^{+}\mu^{-}$ processes of B meson decays. For $\Lambda_{b} \rightarrow$ $\Lambda \mu^{+} \mu^{-}$ decay, their quark level transitions are also $b \rightarrow$ $s \mu^{+}\mu^{-}$, therefore, in this paper, we try to examine the effects of scalar leptoquark models on some observables of $\Lambda_{b} \rightarrow$ $\Lambda \mu^{+} \mu^{-}$ decay, such as the differential decay width, the longitudinal polarization of the dilepton system, the lepton-side forward-backward asymmetry, the baryon-side forward-backward asymmetry and double lepton polarization asymmetries.

The outline of this paper is as follows. In the next section we present the SM theoretical framework for $\Lambda_{b} \rightarrow$ $\Lambda \mu^{+} \mu^{-}$ transitions. In Sec. 3, we introduce the employed scalar leptoquark models , the transition form factors are given in Sec. 4, In Sec. 5, we present the physical observables and numerical analyses. Finally, we will have a concluding section.

\section{\label{sec:level2} $\Lambda_{b} \rightarrow$ $\Lambda \ell^{+} \ell^{-}$ transitions}

At quark level, the rare decay $\Lambda_{b} \rightarrow$ $\Lambda \ell^{+} \ell^{-}$ are governed by the $b \to s \ell^+\ell^-$ transition, its effective Hamiltonian
in the SM can be written as

\begin{eqnarray}\label{eq:Ampfree}
{\cal H}^{eff} &=&
\frac{G_{F} \alpha_{em} V_{tb}V^{*}_{ts}}{2\sqrt{2 \pi}}\bigg[C_{9}^{\rm{eff}}\bar{s}\gamma_{\mu}(1-\gamma_{5})b\bar{\ell}\gamma^{\mu}\ell\nonumber\\
&&+ C_{10}\bar{s}\gamma_{\mu}(1-\gamma_{5})b\bar{\ell}\gamma^{\mu}\gamma_{5}\ell \nonumber\\
&&- 2m_{b}C_{7}^{\rm{eff}}\frac{1}{q^{2}}\bar{s}i\sigma_{\mu\nu}q^{\nu}(1+\gamma_{5})b\bar{\ell}\gamma^{\mu}\ell\bigg],
\end{eqnarray}

where $G_{F}$ is the Fermi constant, $\alpha_{em}=\frac{e^2}{4 \pi}$ is the fine-structure constant and $V_{qq\prime}$ denote the CKM matrix elements.

Following Ref.\cite{william}, the effective Wilson coefficients in the high $q^{2}$ region are given by
\begin{eqnarray}\label{eq:effWC}
 C_{7}^{\rm{eff}}(q^{2})&=& C_{7}- \frac{1}{3}(C_{3}+\frac{4}{3}C_{4}+20C_{5}+\frac{80}{3}C_{6})\nonumber\\
 &&-\frac{\alpha_{s}}{4 \pi}[(C_{1}-6C_{2})F^{(7)}_{1,c}(q^{2})+C_{8}F^{(7)}_{8}(q^{2})],\nonumber
\end{eqnarray}
\begin{eqnarray}
 C_{9}^{\rm{eff}}(q^{2})&=& C_{9}+\frac{4}{3}C_{3}+\frac{64}{9}C_{5}+\frac{64}{27}C_{6}\nonumber\\
 &&+h(0,q^{2})(-\frac{1}{2}C_{3}-\frac{2}{3}C_{4}-8C_{5}-\frac{32}{3}C_{6})\nonumber\\
     &&+h(m_{b},q^{2})(-\frac{7}{2}C_{3}-\frac{2}{3}C_{4}-38C_{5}-\frac{32}{3}C_{6})\nonumber\\
     &&+h(m_{c},q^{2})(\frac{4}{3}C_{1}+C_{2}+6C_{3}+60C_{5})\nonumber\\
     &&-\frac{\alpha_{s}}{4 \pi}[C_{1}F^{(9)}_{1,c}(q^{2})+C_{2}F^{(9)}_{2,c}(q^{2})\nonumber\\
     &&+C_{8}F^{(9)}_{8}(q^{2})],
\end{eqnarray}
where the explicit expressions of these functions $F^{(7,9)}_{8}(q^{2})$, $h(m_{q},q^{2})$, $F^{(7,9)}_{1,c}(q^{2})$ and $F^{(7,9)}_{2,c}(q^{2})$ can be found in Refs.\cite{ddu,dseidel,tfeldmann}. However, in the low $q^{2}$ region, non-factorizable hadronic effects are expected to have the sizeable corrections,
these have not been calculated for the baryonic decay\cite{tfeldmann,yumingwang}.
According to Ref.\cite{william}, we use the effective Wilson coefficients $C_{7}^{\rm{eff}}(q^{2})$ and $C_{9}^{\rm{eff}}(q^{2})$
in Eq.(\ref{eq:effWC}) both in the low $q^{2}$ region and in the high $q^{2}$ region by increasing the $5\%$ uncertainty.

\section{\label{sec:level3} Scalar leptoquark models}

Here we consider two kinds of the minimal renormalizable scalar leptoquark models\cite{nejc1},
containing one single additional representation of $SU(3)\times SU(2)\times U(1)$ where baryon number violation can not be allowed in perturbation theory. There are only two such models which are represented as $X(3,2,7/6)$ and $X(3,2,1/6)$ under the $SU(3)\times SU(2)\times U(1)$ gauge group.

The interaction Lagrangian for the scalar leptoquark $X(3,2,7/6)$ couplings to the fermion bilinear can be written as
\begin{eqnarray}
{\cal L}= -\lambda^{ij}_{u}\bar{u}^{i}_{R}X^{T}\epsilon L^{j}_{L}-\lambda^{ij}_{e}\bar{e}^{i}_{R}X^{\dagger}Q^{j}_{L}+h.c.,
\end{eqnarray}
where $i,j=1,2,3$ are the generation indices, the couplings $\lambda$ are in general complex parameters, $u_{R}(e_{R})$ is the right handed up-type quark(charged lepton) singlet, $X$ is the scalar leptoquark doublet, $\epsilon=i\sigma_{2}$ is a $2\times2$ matrix and $Q_{L}(L_{L})$ is the left handed quark (lepton) doublet.

After performing Fierz transformation, the contribution to the interaction Hamiltonian for the $b\rightarrow s \mu^{+}\mu^{-}$ is
\begin{eqnarray}
\mathcal{H}&=&\frac{\lambda^{32}_{\mu}\lambda^{22*}_{\mu}}{8M^{2}_{(7/6)}}[\bar{s}\gamma^{\mu}(1-\gamma_{5})b][\bar{\mu}\gamma_{\mu}(1+\gamma_{5})\mu]\nonumber\\
&=&\frac{\lambda^{32}_{\mu}\lambda^{22*}_{\mu}}{4M^{2}_{(7/6)}}(O_{9}+O_{10}),
\end{eqnarray}
which can be written in the style of the SM effective Hamiltonian as
\begin{eqnarray}
\mathcal{H}=-\frac{G_{F}\alpha}{\sqrt{2} \pi}V_{tb}V^{*}_{ts}(C^{NP}_{9}O_{9}+C^{NP}_{10}O_{10}),
\end{eqnarray}
Then we obtain the new Wilson coefficients
\begin{eqnarray}
C^{NP}_{9}=C^{NP}_{10}=-\frac{\pi}{2\sqrt{2}G_{F} V_{tb}V^{*}_{ts}\alpha}\frac{\lambda^{32}_{\mu}\lambda^{22*}_{\mu}}{M^{2}_{(7/6)}},
\end{eqnarray}

The interaction Lagrangian for the scalar leptoquark $X=(3,2,1/6)$ couplings to the fermion bilinear can be written as
\begin{eqnarray}
{\cal L}= -\lambda^{ij}_{d}\bar{d}^{i}_{R}X^{T}\epsilon L^{j}_{L}+h.c.,
\end{eqnarray}

After performing Fierz transformation, the contribution to the interaction Hamiltonian for the $b\rightarrow s \mu^{+}\mu^{-}$ is
\begin{eqnarray}
\mathcal{H}&=&\frac{\lambda^{22}_{s}\lambda^{32*}_{b}}{8M^{2}_{(1/6)}}[\bar{s}\gamma^{\mu}(1+\gamma_{5})b][\bar{\mu}\gamma_{\mu}(1-\gamma_{5})\mu]\nonumber\\
&=&\frac{\lambda^{22}_{s}\lambda^{32*}_{b}}{4M^{2}_{(1/6)}}(O^{'}_{9}-O^{'}_{10}),
\end{eqnarray}
where $O^{'}_{9}$ and $O^{'}_{10}$ are dimension-six operators obtained from $O_{9}$ and $O_{10}$ by the replacement $P_{L}\leftrightarrow P_{R}$. Writing in the style of the SM effective Hamiltonian, then we obtain the new Wilson coefficients
\begin{eqnarray}
C^{'NP}_{9}=-C^{'NP}_{10}=\frac{\pi}{2\sqrt{2}G_{F} V_{tb}V^{*}_{ts}\alpha}\frac{\lambda^{22}_{s}\lambda^{32*}_{b}}{M^{2}_{(1/6)}},
\end{eqnarray}

In Ref.\cite{suchismita2}, comparing the bounds on NP coupling parameters obtained from $B_{s}\rightarrow \mu^{+}\mu^{-}$, $\bar{B}^{0}_{d}\rightarrow X_{s}\mu^{+}\mu^{-}$ and $B_{s}-\bar{B}_{s}$ mixing, respectively, the authors
obtain the following results
\begin{widetext}
\begin{eqnarray}
0\leq \frac{|\lambda^{32}_{\mu}\lambda^{22*}_{\mu}|}{M^{2}_{(7/6)}}=\frac{|\lambda^{22}_{s}\lambda^{32*}_{b}|}{M^{2}_{(1/6)}}=\frac{|\lambda^{32}_{s}\lambda^{22*}_{b}|}{M^{2}_{S}} \leq 5\times 10^{-9}~~\rm{GeV}^{-2},~~~~for~~\pi/2\leq \phi^{NP} \leq 3\pi/2.
\end{eqnarray}
\end{widetext}
where the bounds will be used in the process of our calculations.

\section{\label{sec:level4} Transition form factors}

For $\Lambda_{b} \rightarrow \Lambda \mu^{+}\mu^{-}$ decay, these form factors have been calculated in the framework of QCD light-cone sum rules (LCSR) in the low $q^{2}$ region\cite{yumingwang} and lattice QCD in the high $q^{2}$ region \cite{william}, respectively. All of them use the helicity-based definition of the form factors\cite{mwyip}.
\begin{widetext}
\begin{eqnarray}\label{factor}
<\Lambda(p^{\prime},s^{\prime})|\bar{s}\gamma^{\mu}b|\Lambda_{b}(p,s)>&=&\bar{u}_{\Lambda}(p^{\prime},s^{\prime})[f_{0}(q^{2})(m_{\Lambda_{b}}-m_{\Lambda})\frac{q^{\mu}}{q^{2}}\nonumber\\
&&+f_{+}(q^{2})\frac{m_{\Lambda_{b}}+m_{\Lambda}}{s_{+}}(p^{\mu}+p^{\prime \mu}-(m_{\Lambda_{b}}^{2}-m_{\Lambda}^{2})\frac{q^{\mu}}{q^{2}})\nonumber\\
&&+f_{\perp}(q^{2})(\gamma^{\mu}-\frac{2m_{\Lambda}}{s_{+}}p^{\mu}-\frac{2m_{\Lambda_{b}}}{s_{+}}p^{\prime \mu})]u_{\Lambda_{b}}(p,s),\nonumber\\
<\Lambda(p^{\prime},s^{\prime})|\bar{s}\gamma^{\mu}\gamma_{5}b|\Lambda_{b}(p,s)>&=&-\bar{u}_{\Lambda}(p^{\prime},s^{\prime})\gamma_{5}[g_{0}(q^{2})(m_{\Lambda_{b}}+m_{\Lambda})\frac{q^{\mu}}{q^{2}}\nonumber\\
&&+g_{+}(q^{2})\frac{m_{\Lambda_{b}}-m_{\Lambda}}{s_{-}}(p^{\mu}+p^{\prime \mu}-(m_{\Lambda_{b}}^{2}-m_{\Lambda}^{2})\frac{q^{\mu}}{q^{2}})\nonumber\\
&&+g_{\perp}(q^{2})(\gamma^{\mu}+\frac{2m_{\Lambda}}{s_{-}}p^{\mu}-\frac{2m_{\Lambda_{b}}}{s_{-}}p^{\prime \mu})]u_{\Lambda_{b}}(p,s),\nonumber\\
<\Lambda(p^{\prime},s^{\prime})|\bar{s}i\sigma^{\mu \nu}q_{\nu}b|\Lambda_{b}(p,s)>&=&-\bar{u}_{\Lambda}(p^{\prime},s^{\prime})[h_{+}(q^{2})\frac{q^{2}}{s_{+}}(p^{\mu}+p^{\prime \mu}-(m_{\Lambda_{b}}^{2}-m_{\Lambda}^{2})\frac{q^{\mu}}{q^{2}})\nonumber\\
&&+h_{\perp}(q^{2})(m_{\Lambda_{b}}+m_{\Lambda})(\gamma^{\mu}-\frac{2m_{\Lambda}}{s_{+}}p^{\mu}-\frac{2m_{\Lambda_{b}}}{s_{+}}p^{\prime \mu})]u_{\Lambda_{b}}(p,s),\nonumber\\
<\Lambda(p^{\prime},s^{\prime})|\bar{s}i\sigma^{\mu \nu}q_{\nu}\gamma_{5}b|\Lambda_{b}(p,s)>&=&-\bar{u}_{\Lambda}(p^{\prime},s^{\prime})\gamma_{5}[{\tilde{h}}_{+}(q^{2})\frac{q^{2}}{s_{-}}(p^{\mu}+p^{\prime \mu}-(m_{\Lambda_{b}}^{2}-m_{\Lambda}^{2})\frac{q^{\mu}}{q^{2}})\nonumber\\
&&+\tilde{h}_{\perp}(q^{2})(m_{\Lambda_{b}}-m_{\Lambda})(\gamma^{\mu}+\frac{2m_{\Lambda}}{s_{-}}p^{\mu}-\frac{2m_{\Lambda_{b}}}{s_{-}}p^{\prime \mu})]u_{\Lambda_{b}}(p,s),
\end{eqnarray}
where $q=p-p^{\prime}$ and $s_{\pm}=(m_{\Lambda_{b}}\pm m_{\Lambda})^{2}-q^{2}$. The fit functions of helicity-based form factors can be found in Eqs.(133)-(135) of Ref.\cite{yumingwang} and Eq.(49) of Ref.\cite{william}.
\end{widetext}

\section{\label{sec:level5}Physical observables and numerical analyses}
\subsection*{\small V.1 Some measured observables}
According to Ref.\cite{aaij1:2013st}, the $\Lambda_{b}$ polarization at the LHC has been measured to be small and compatible with zero, and polarization effects will average out for the symmetric ATLAS and CMS detectors, so we consider the initial baryon $\Lambda_{b}$ as unpolarized. The four-fold differential rate of the $\Lambda_{b} \rightarrow \Lambda(\rightarrow a(1/2^{+}), b(0^{-})) \ell^{+}\ell^{-}$ can be written as\cite{philipp}
\begin{widetext}

\begin{eqnarray}
\frac{d^{4}\Gamma}{dq^{2}d\cos\theta_{\ell}d\cos\theta_{\Lambda}d\phi}&=&\frac{3}{8 \pi}[(K_{1ss}\sin^{2}\theta_{\ell}+K_{1cc}\cos^{2}\theta_{\ell}+K_{1c}\cos\theta_{\ell})+(K_{2ss}\sin^{2}\theta_{\ell}+K_{2cc}\cos^{2}\theta_{\ell}+K_{2c}\cos\theta_{\ell})\cos\theta_{\Lambda}\nonumber\\
&&+(K_{3sc}\sin\theta_{\ell}\cos\theta_{\ell}+K_{3s}\sin\theta_{\ell})\sin\theta_{\Lambda}\sin\phi+(K_{4sc}\sin\theta_{\ell}\cos\theta_{\ell}+K_{4s}\sin\theta_{\ell})\sin\theta_{\Lambda}\sin\phi].
\end{eqnarray}
\end{widetext}
where the angles $\theta_{\ell}$ and $\theta_{\Lambda}$ denote the polar directions of $\ell^{-}$ and $a(1/2^{+})$, respectively. $\phi$ is the azimuthal angle between the $\ell^{+}\ell^{-}$ and $a(1/2^{+})b(0^{-})$ decay planes, and the explicit expressions of the coefficients $K_{i}$ can be found in Ref.\cite{philipp}.

\noindent (a) The differential decay width
\begin{equation}
\frac{d\Gamma}{dq^{2}}=2K_{1ss}+K_{1cc},
\end{equation}
(b) The longitudinal polarization of the dilepton system
\begin{equation}
F_{L}=\frac{2K_{1ss}-K_{1cc}}{2K_{1ss}+K_{1cc}},
\end{equation}
(c) The lepton-side forward-backward asymmetry
\begin{equation}
A_{FB}^{\ell}=\frac{3}{2}\frac{K_{1c}}{2K_{1ss}+K_{1cc}},
\end{equation}
(d) The baryon-side forward-backward asymmetry
\begin{equation}
A_{FB}^{\Lambda}=\frac{1}{2}\frac{2K_{2ss}+K_{2cc}}{2K_{1ss}+K_{1cc}},
\end{equation}

In the process of numerical analyses, we consider the theoretical uncertainties of all input parameters. For the form factors, we use the results of QCD light-cone sum rules (LCSR) in the low $q^{2}$ region\cite{yumingwang} and lattice QCD in the high $q^{2}$ region\cite{william}. Comparing to the current data which have been measured by LHCb collaboration\cite{LHCB}, we plot the dependence of four observables mentioned above on the full physical region except the intermediate region of $q^{2}$ in Fig. 1.

From Fig. 1, we obtain the following results:
\begin{itemize}

\item For the differential decay width $\frac{d\Gamma}{dq^{2}}$, its prediction values of SM are consistent with the current data in the ranges of $0.1<q^{2}<1 \rm{GeV}^2/c^{2}$ and $15<q^{2}<16 \rm{GeV}^2/c^{2}$. When we consider the effects of this two NP models, the theoretical predictions are still consistent with the experimental results with $1\sigma$ in these ranges. However, in the remaining ranges, its prediction values of SM and this two NP models are difficult to meet the current data. But in the large $q^{2}$ region, the prediction values of the scalar leptoquark $X(3,2,7/6)$ are more closer to the current data.

\item For the longitudinal polarization $F_{L}$ of the dilepton system, its prediction values of SM and this two NP models are consistent with the current data both in the low $q^{2}$ region and in the high $q^{2}$ region, respectively. In the low $q^{2}$ region, the prediction value of the scalar leptoquark $X(3,2,7/6)$ enhances that of SM, but the opposite result happens in the scalar leptoquark $X(3,2,1/6)$. There are not obvious difference results between the SM and this two NP models in the high $q^{2}$ region.

\item For the lepton-side forward-backward asymmetry $A_{FB}^{\mu}$, in the range of $0.1<q^{2}<1 \rm{GeV}^2/c^{2}$, its prediction value of SM is consistent with the current data with $1\sigma$, but the result of the scalar leptoquark $X(3,2,7/6)$ is more closer to the central value of the current data than that of SM. In the high $q^{2}$ region, its prediction value of SM is lower than the current data. But the result of the scalar leptoquark $X(3,2,7/6)$ can meet the current data in the range of $15<q^{2}<16 \rm{GeV}^2/c^{2}$.

\item For the baryon-side forward-backward asymmetry $A_{FB}^{\Lambda}$, except in the range of $16<q^{2}<18 \rm{GeV}^2/c^{2}$, the current data in the remaining ranges can be met both in the SM and in this two NP models, respectively. When we consider the NP effects, this observable show strong dependence on the scalar leptoquark $X(3,2,1/6)$. However, there are not obvious difference results between the SM and scalar leptoquark $X(3,2,7/6)$.

\end{itemize}

\subsection*{\small V.2  Double lepton polarization asymmetries}

The definition of the double lepton polarization asymmetry can be written as
\begin{widetext}
\begin{eqnarray} \LARGE{ \mathcal{P}_{ij}=\frac{(\frac{d\Gamma(\vec{s}_{i}^{-},\vec{s}_{j}^{+})}{d\hat{s}}
-\frac{d\Gamma(-\vec{s}_{i}^{-},\vec{s}_{j}^{+})}{d\hat{s}})
-(\frac{d\Gamma(\vec{s}_{i}^{-},-\vec{s}_{j}^{+})}{d\hat{s}}
-\frac{d\Gamma(-\vec{s}_{i}^{-},-\vec{s}_{j}^{+})}{d\hat{s}})}
{(\frac{d\Gamma(\vec{s}_{i}^{-},\vec{s}_{j}^{+})}{d\hat{s}}
+\frac{d\Gamma(-\vec{s}_{i}^{-},\vec{s}_{j}^{+})}{d\hat{s}})
+(\frac{d\Gamma(\vec{s}_{i}^{-},-\vec{s}_{j}^{+})}{d\hat{s}}
+\frac{d\Gamma(-\vec{s}_{i}^{-},-\vec{s}_{j}^{+})}{d\hat{s}})},}
\end{eqnarray}
\end{widetext}
where $\vec{s}_{i(j)}^{-(+)}$ is the orthogonal unit vector in the rest frame of the leptons, its explicit explanation and nine double-lepton polarization asymmetries are presented in\cite{tmaliev2}.

In Ref.\cite{tmaliev2}, the form factors are defined as follows
\begin{widetext}
\begin{eqnarray}\label{factor1}
\langle \Lambda(p)|\bar{s} \gamma_{\mu} (1-\gamma_{5}) b|\Lambda_{b}(p+q)\rangle&=&\bar{u}_{\Lambda}(p)\Big[\gamma_{\mu}f_{1}(q^2)+i\sigma_{\mu\nu}q^{\nu}f_{2}(q^2)+q^{\mu}f_{3}(q^2)\nonumber\\
&&-\gamma_{\mu}\gamma_{5}g_{1}(q^2)-i\sigma_{\mu\nu}\gamma_{5}q^{\nu}g_{2}(q^2)-q^{\mu}\gamma_{5}g_{3}(q^2)\Big]u_{\Lambda_{b}}(p+q),\nonumber\\
\langle \Lambda(p)|\bar{s}i\sigma_{\mu\nu}q^{\nu}(1+\gamma_{5}) b|\Lambda_{b}(p+q)\rangle&=&\bar{u}_{\Lambda}(p)\Big[\gamma_{\mu}f_{1}^{T}(q^2)+i\sigma_{\mu\nu}q^{\nu}f_{2}^{T}(q^2)+q^{\mu}f_{3}^{T}(q^2)\nonumber\\
&&+\gamma_{\mu}\gamma_{5}g_{1}^{T}(q^2)+i\sigma_{\mu\nu}\gamma_{5}q^{\nu}g_{2}^{T}(q^2)+q^{\mu}\gamma_{5}g_{3}^{T}(q^2)\Big]u_{\Lambda_{b}}(p+q),
\end{eqnarray}
\end{widetext}
where $u_{\Lambda_{b}}$ and $u_{\Lambda}$ are spinors of $\Lambda_{b}$ and $\Lambda$ baryons, respectively.

The form factors $f_{i}^{(T)}$ and $g_{i}^{(T)}$ in Eq.(\ref{factor1}) are related to the helicity form factors $f_{+,\bot,0}$, $g_{+,\bot,0}$, $h_{+,\bot}$ and $\tilde{h}_{+,\bot}$ in Eq.(\ref{factor}) as follows
\begin{eqnarray}
f_{+}&=&f_{1}-\frac{q^{2}}{m_{\Lambda_{b}}+m_{\Lambda}}f_{2},\nonumber \\f_{\bot}&=&f_{1}-(m_{\Lambda_{b}}+m_{\Lambda})f_{2},\nonumber \\f_{0}&=&f_{1}+\frac{q^{2}}{m_{\Lambda_{b}}-m_{\Lambda}}f_{3},\nonumber \\g_{+}&=&g_{1}+\frac{q^{2}}{m_{\Lambda_{b}}-m_{\Lambda}}g_{2},\nonumber \\
g_{\bot}&=&g_{1}+(m_{\Lambda_{b}}-m_{\Lambda})g_{2},\nonumber 
\end{eqnarray}
\begin{eqnarray}
g_{0}&=&g_{1}-\frac{q^{2}}{m_{\Lambda_{b}}+m_{\Lambda}}g_{3},\nonumber \\h_{+}&=&f_{2}^{T}-\frac{m_{\Lambda_{b}}+m_{\Lambda}}{q^{2}}f_{1}^{T},\nonumber\\h_{\bot}&=&f_{2}^{T}-\frac{f_{1}^{T}}{m_{\Lambda_{b}}+m_{\Lambda}},\nonumber \\
\tilde{h}_{+}&=&g_{2}^{T}+\frac{m_{\Lambda_{b}}-m_{\Lambda}}{q^{2}}g_{1}^{T},\nonumber \\ \tilde{h}_{\bot}&=&g_{2}^{T}+\frac{g_{1}^{T}}{m_{\Lambda_{b}}-m_{\Lambda}},
\end{eqnarray}
The amplitude for $\Lambda_{b}\to \Lambda \mu^{+}\mu^{-}$ decay can be written in terms of twelve form factors in Eq.(\ref{factor1}), and we
find that
\begin{widetext}

\begin{eqnarray}
\mathcal{M}&=&\frac{G\alpha}{8\sqrt{2} \pi}V_{tb}V^{*}_{ts}\biggl\{\bar{\ell}\gamma^{\mu}(1-\gamma_{5})\ell\bar{u}_{\Lambda}(p)\{(A_{1}-D_{1})
\gamma_{\mu}(1+\gamma_{5})+(B_{1}+E_{1})\gamma_{\mu}(1-\gamma_{5})\nonumber\\&&
+i\sigma_{\mu\nu}q^{\nu}[(A_{2}-D_{2})(1+\gamma_{5})+(B_{2}-E_{2})(1-\gamma_{5})]\nonumber\\&&
+q_{\mu}[(A_{3}-D_{3})(1+\gamma_{5})+(B_{3}-E_{3})(1-\gamma_{5})]\}u_{\Lambda_{b}}(p+q)\nonumber\\&&
+\bar{\ell}\gamma_{\mu}(1+\gamma_{5})\ell\bar{u}_{\Lambda}(P)\{(A_{1}+D_{1})\gamma_{\mu}(1+\gamma_{5})+(B_{1}+E_{1})\gamma_{\mu}(1-\gamma_{5})\nonumber\\&&
+i\sigma_{\mu\nu}q^{\nu}[(A_{2}+D_{2})(1+\gamma_{5})+(B_{2}+E_{2})(1-\gamma_{5})]\nonumber\\&&
+q_{\mu}[(A_{3}+D_{3})(1+\gamma_{5})+(B_{3}+E_{3})(1-\gamma_{5})]\}u_{\Lambda_{b}}(p+q)\biggl\},
\end{eqnarray}
\end{widetext}
where
\begin{eqnarray}
A_{1}&=&-\frac{2m_{b}}{q^{2}}C^{eff}_{7}(f^{T}_{1}+g^{T}_{1})+(C^{eff}_{9}+C^{NP}_{9})(f_{1}-g_{1})\nonumber\\
&&+C^{'NP}_{9}(f_{1}+g_{1}),\nonumber\\
A_{2}&=&A_{1}(1\rightarrow 2),\nonumber\\
A_{3}&=&A_{1}(1\rightarrow 3),\nonumber\\
B_{i}&=&A_{i}(g_{i}\rightarrow -g_{i};g_{i}^{T}\rightarrow -g_{i}^{T}),\nonumber\\
D_{1}&=&(C^{eff}_{10}+C^{NP}_{10})(f_{1}-g_{1})+C^{'NP}_{10}(f_{1}+g_{1}),\nonumber\\
D_{2}&=&D_{1}(1\rightarrow 2),\nonumber\\
D_{3}&=&D_{1}(1\rightarrow 3),\nonumber\\
E_{i}&=&D_{i}(g_{i}\rightarrow -g_{i}),
\end{eqnarray}
Because the vector and axial vector currents are not renormalized, so we neglect the anomalous dimensions of coefficients $C^{(')NP}_{9}$ and $C^{(')NP}_{10}$\cite{bobeth4}.

We also plot the dependence of the double lepton polarization asymmetries on the full physical region except the intermediate region of $q^{2}$ in Fig. 2, and find,

\begin{itemize}

\item Double lepton polarization asymmetries $P_{LT}$, $P_{TL}$, $P_{NN}$ and $P_{TT}$ of this
 decay process are sensitive to the contribution of the scalar leptoquark $X(3,2,7/6)$
but not to that of the scalar leptoquark $X(3,2,1/6)$.

\item For double lepton polarization asymmetry $P_{LT}$ of $\Lambda_{b}\to \Lambda \mu^{+}\mu^{-}$ decay,
the contribution of the scalar leptoquark $X(3,2,7/6)$ can enhance its maximum value of SM prediction from 0.48 to 0.65 in the low $q^{2}$ region. For this decay process, this effects of this two NP models on $P_{TL}$ which is not presented in this paper are similar to $P_{LT}$, respectively.

\item For double lepton polarization asymmetry $P_{NN}$ of $\Lambda_{b}\to \Lambda \mu^{+}\mu^{-}$ decay,
when we consider the contribution of the scalar leptoquark $X(3,2,7/6)$, its value of SM prediction can be enhanced quite a lot both in the low $q^{2}$ region and in the high $q^{2}$ region. For this decay process, the contribution of the scalar leptoquark $X(3,2,7/6)$ to $P_{TT}$ is similar to $P_{NN}$, respectively.

\item For double lepton polarization asymmetries $P_{LN}$, $P_{NL}$, $P_{NT}$ and $P_{TN}$, their values of SM prediction are almost zero, and the effects of this two NP models are not significant on them.

\end{itemize}

\section{\label{sec:level6}Conclusions}
We calculate the differential decay width, the longitudinal polarization of the dilepton system, the lepton-side forward-backward asymmetry, the baryon-side forward-backward asymmetry and double lepton polarization asymmetries of $\Lambda_{b} \rightarrow$ $\Lambda \mu^{+} \mu^{-}$ decay in the scalar leptoquark model $X(3,2,7/6)$ and $X(3,2,1/6)$, respectively. Using the constrained parameter spaces from $B_{s} \rightarrow \mu^{+}\mu^{-}$ and $B_{d} \rightarrow X_{s}\mu^{+}\mu^{-}$ decays, we depict the correlative figures between these observables and the momentum transfer $q^{2}$, respectively. We find, for the differential decay width, the longitudinal polarization of the dilepton system, the lepton-side forward-backward asymmetry and the baryon-side forward-backward asymmetry, which have been measured by LHCb collaboration, most of their current data can be met both in the SM and in this two NP models. However, some of their current data can still not be met in the SM. When we consider the effects of this two NP models, like the lepton-side forward-backward asymmetry $A_{FB}^{\mu}$ in the range of $0.1<q^{2}<1 \rm{GeV}^2/c^{2}$, its current data with $1\sigma$ can be met. For the double lepton polarization asymmetries, $P_{LT}$, $P_{TL}$, $P_{NN}$ and $P_{TT}$ show strong dependence on the scalar leptoquark model $X(3,2,7/6)$ but not on $X(3,2,1/6)$, respectively. However, the prediction values of $P_{LN}$, $P_{NL}$, $P_{TN}$ and $P_{NT}$ in the SM are almost zero, and they also show weak dependence on this two NP models.

\section*{Acknowledgments}
The work is supported by the National Science Foundation under contract Nos.11225523 and U1332103.

\begin{figure}
\includegraphics[width=7cm,height=5cm]{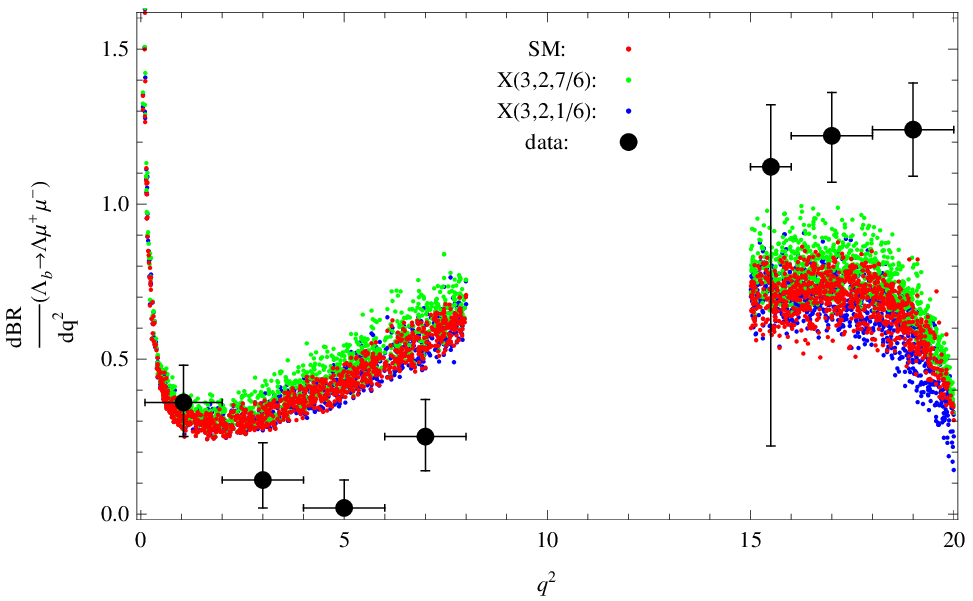}
\includegraphics[width=7cm,height=5cm]{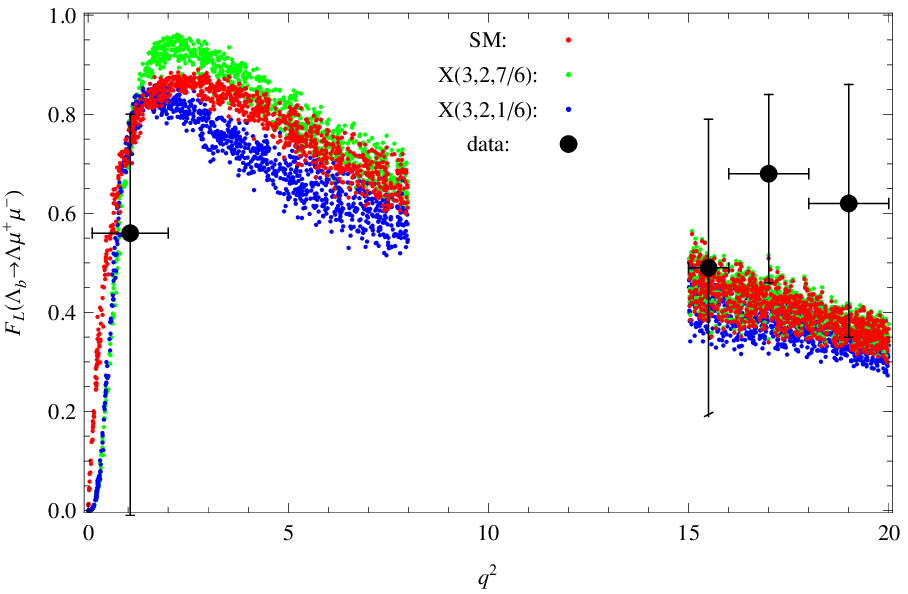}
\includegraphics[width=7cm,height=5cm]{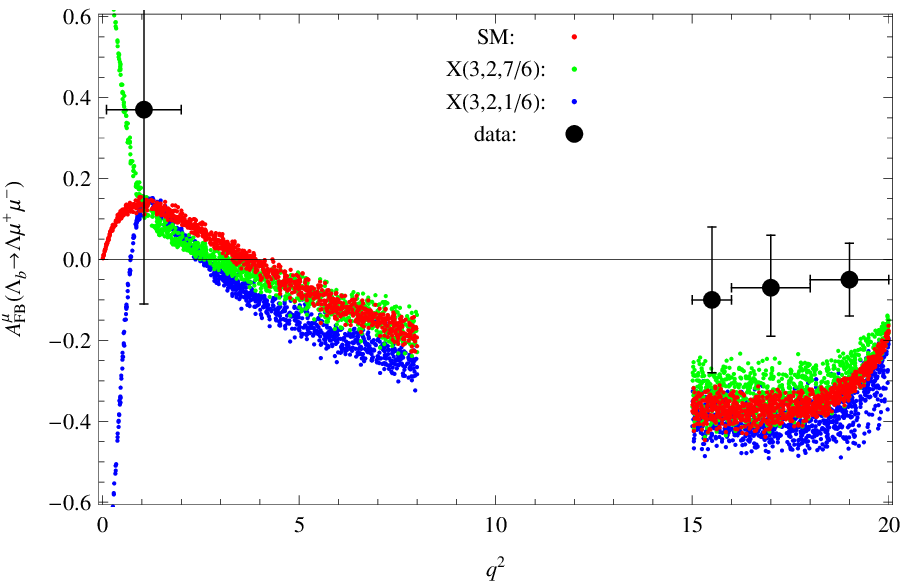}
\includegraphics[width=7cm,height=5cm]{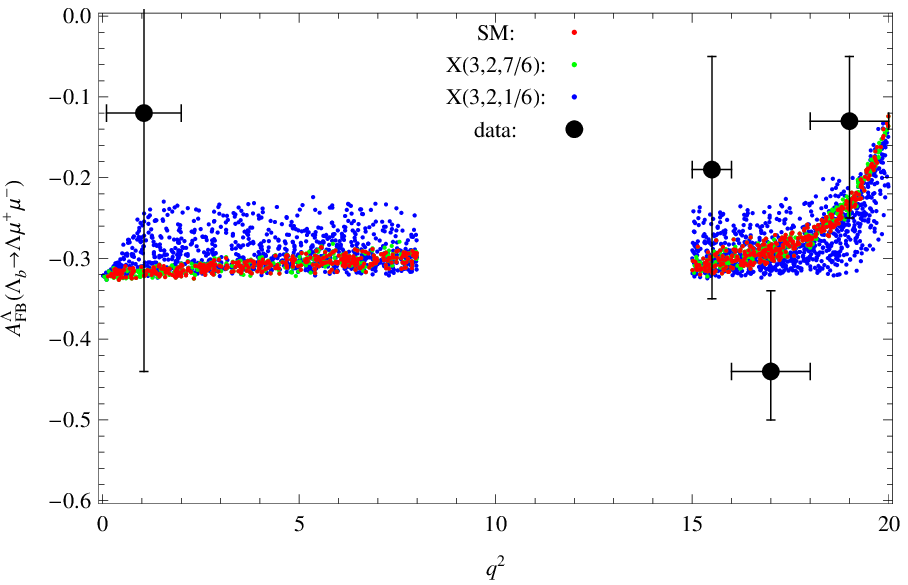}
\caption{The dependence of $\frac{d\Gamma}{dq^{2}}$, $F_{L}$, $A_{FB}^{\mu}$ and $A_{FB}^{\Lambda}$ on $q^{2}$, respectively.}
\end{figure}
\begin{figure}
\includegraphics[width=7cm,height=5cm]{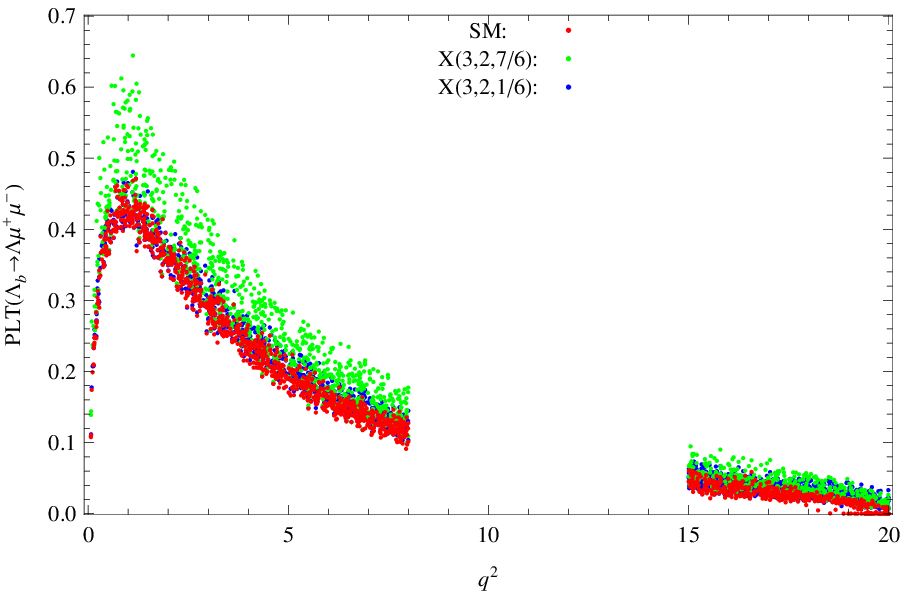}
\includegraphics[width=7cm,height=5cm]{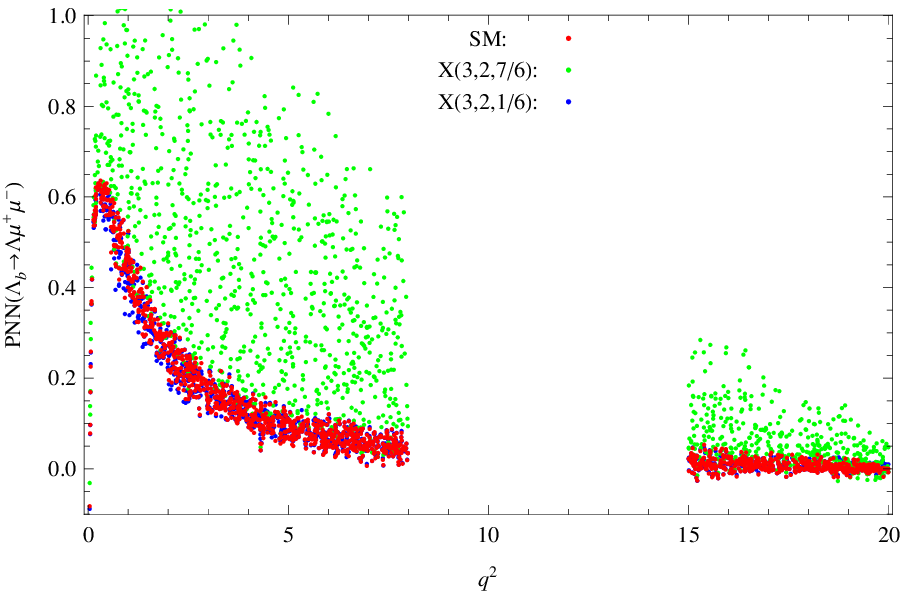}
\includegraphics[width=7cm,height=5cm]{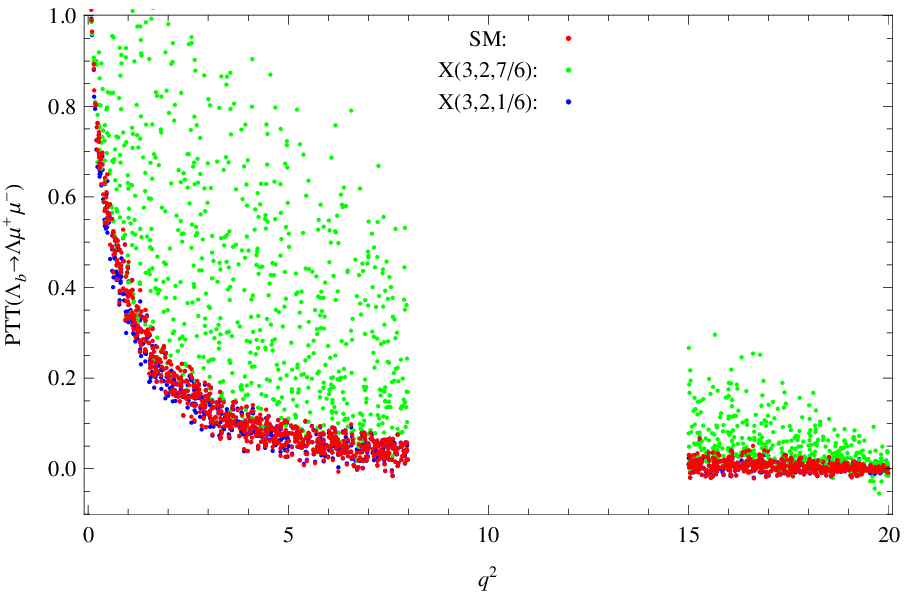}
\caption{The dependence of double lepton polarization asymmetries $P_{LT}$, $P_{NN}$ and $P_{TT}$ on the $q^{2}$, respectively.}
\end{figure}

\end{document}